# Attitude Control of Solar Sail with Reflectivity Control Devices

Authors: Pierce S. Boughton & Yang Yang

## Abstract
Solar sails offer a promising solution for fuel free propulsion, enabling novel mission profiles and deeper space exploration. While reaction wheels are standard for spacecraft attitude control, the large moment of inertia of solar sails often lead to frequent reaction wheel saturation, necessitating momentum offloading via additional control methods. Magnetorquers have historically been used for this purpose. This paper investigates Reflectivity Control Devices (RCDs) as an alternative method for momentum management, aiming to prevent reaction wheel saturation. A dynamic model of a solar sail in a Sun synchronous orbit is developed, incorporating disturbance torques to assess control. Numerical simulations evaluate the effectiveness of RCDs in offloading reaction wheel momentum and preventing saturation. The results indicate additional applications for RCDs in Earth orbit as well as potential for deep space missions where magnetorquers cannot be used.

Pierce Boughton | pierce.boughton@student.unsw.edu.au[1]
Yang Yang | yang.yang16@unsw.edu.au

School of Mechanical and Manufacturing Engineering, UNSW Sydney

---

[1] Permanent address: pierce.b@live.com.au

# 1  Introduction

Solar Radiation Pressure (SRP) is known as one of the most significant environmental disturbances in space and must be considered in mission design (Starin and Eterno, 2011). Since photons are known to have momentum, a pressure force is exerted upon objects in sunlight as shown in Fig. 1 below. The concept of solar sailing is to utilise the SRP as a source of propulsion for orbital manoeuvres and attitude adjustment in place of or alongside typical propulsion systems typically used by spacecraft like thrusters. To maximise the resulting force, surface area needs to be maximised while mass is minimised. The sail is initially stowed to reduce the spacecraft size during launch. Once in space, a solar sail consisting of a thin, reflective sheet is deployed, increasing the surface area of the spacecraft and maximising SRP.

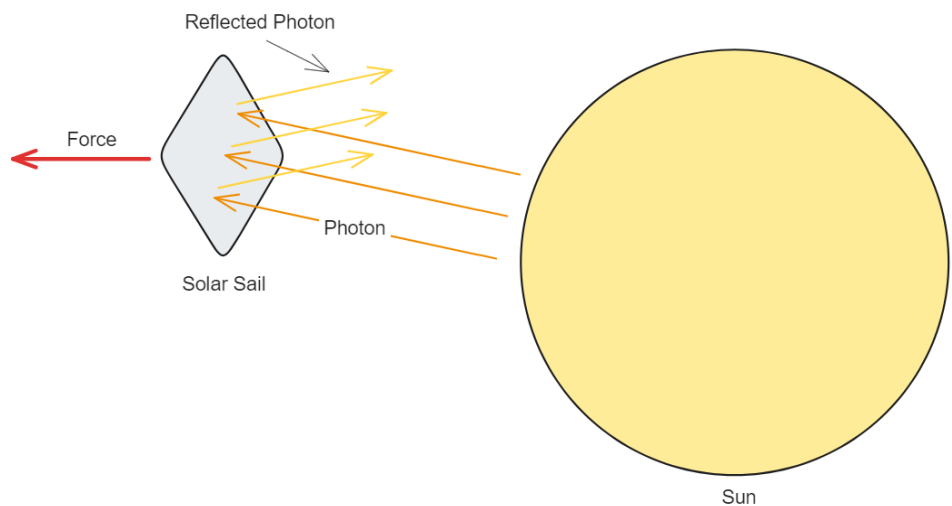

*Figure 1. Solar sailing concept using solar radiation pressure for propulsion.*

## 1.1  Background

JAXA's IKAROS mission (Tsuda et al., 2013, Tsuda et al., 2011) was the first successful demonstration of solar sailing. IKAROS used a 200 m² sail attached to a 307 kg spacecraft. The mission employed a spinning sail to increase the area to mass ratio. Crucially, IKAROS featured Reflectivity Control Devices (RCDs), activated in a timed manner during the spin to create an imbalance in SRP, generating the required torques to adjust attitude. Many missions since IKAROS have not considered the use of RCD technology for attitude control, even with its simplicity and little mass addition.

The LightSail 2 mission (Spencer et al., 2021, Mansell et al., 2020) utilised rigid booms to deploy the solar sail. The spacecraft employed a single axis reaction wheel along with magnetorquers to perform rapid slews to maximise SRP at key orbital points. However, LightSail 2 encountered issues with reaction wheel saturation that necessitated frequent momentum dumping using magnetorquers. This limited the efficiency of the orbit raising as precise control could not be continuously maintained. Although effective in demonstrating controlled solar sailing, these challenges illustrate the limits of current attitude control systems.

*1.2 Attitude Control Methods*

Fu et al., 2016 states that conventional attitude control systems are insufficient for use with solar sails due to the increase in disturbance torques. An overview of solar sail attitude control methods is provided including control vanes, gimbaled/sliding masses, and shifted/tilted wings. Many of these methods contain significant complexity and have no flight heritage. The CubeSail mission (Lappas et al., 2011) aimed to demonstrate 3 axis attitude control using a combination of magnetorquers and a two dimensional translation unit that adjusted the centre of mass relative to the centre of pressure. While mass efficient and mechanically simple, the mission suffered a critical communications failure before full validation. Simulations indicated a low pointing accuracy due to variation in the Sun angle, limiting control torques. NASA's Near Earth Asteroid (NEA) Scout (McNutt et al., 2014, Orphee et al., 2017) was another unsuccessful mission. The attitude was to be controlled using reaction wheels with thrusters and a translation unit primarily for offloading momentum from the reaction wheels and preventing saturation. Some attitude control methods from previous missions are summarised in Table 1 below.

*Table 1. Summary of solar sail attitude control methods*

| Mission | Attitude Control Method | Notes |
|---|---|---|
| IKAROS (Tsuda et al., 2013, Tsuda et al., 2011) | RCDs, thrusters | RCDs used for 2 axis control |
| LightSail 2 (Spencer et al., 2021, Mansell et al., 2020) | 1 reaction wheel, 3 magnetorquers | Magnetorquers used for offloading momentum |
| CubeSail (Lappas et al., 2011) | Translation unit, 3 magnetorquers | 2 axis attitude control with translation unit. Limited pointing accuracy |
| NEA Scout (McNutt et al., 2014) | 4 reaction wheels, translation unit, thrusters | Reaction wheels used for attitude control with translation unit and thrusters used for offloading momentum |
| ACS3 (Wilkie and Fernandez, 2023) | 4 reaction wheels, 3 magnetorquers | Limited information currently available |

In previous solar sail missions, reaction wheels have commonly been chosen due to their low mass and pointing accuracy (Orphee et al., 2017). However, the large moment of inertia of the sail provides some challenges with attitude control. Reaction wheels suffer from saturation due to the increase in disturbance torques and additional control requirements (Fu et al., 2016). Additional actuators are therefore required for momentum dumping. RCDs have not been used since the IKAROS mission but have potential for offloading momentum in 2 axes.

*1.2.1 Reflectivity Control Devices*

RCDs are thin flexible membranes, placed on the surface of a solar sail to change the reflective properties of the sail with an applied voltage. RCDs contain liquid crystal, which allow the device to change between specular reflection and diffusive reflection (Tsuda et al., 2013). Since the SRP force is dependent on the reflective properties, by changing the reflectivity state of different areas of the sail, a torque is created, allowing for attitude control of the spacecraft as shown in Fig. 2 below.

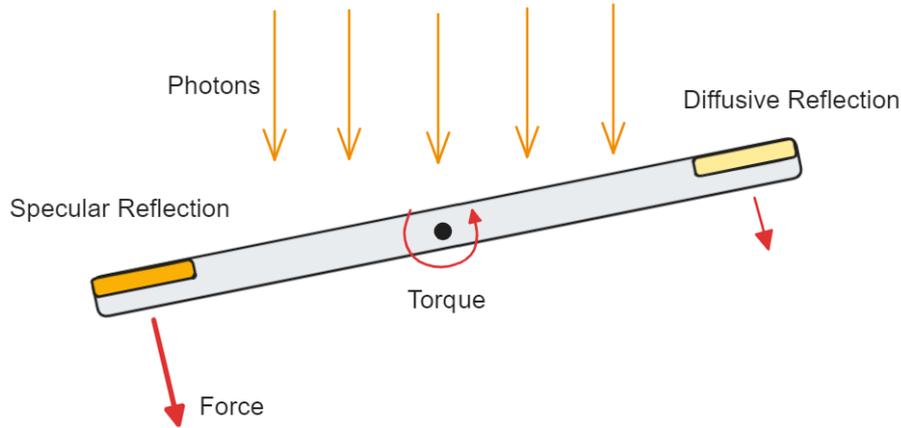

*Figure 2. RCD solar sail torque for attitude control. Specular reflection produces more force than diffusive reflection resulting in a torque being produced.*

RCDs offer a simple and lightweight approach to attitude control for solar sails. They are one of the few successfully demonstrated attitude control devices for solar sails and have only been previously considered for non-rigid spinning solar sails. With advancing structural boom technology (Friesen, 2024), higher area to mass ratios are being achieved with rigid sails, leading to larger adoption of rigid sails over spinning sails (Ancona and Kezerashvili, 2024). Although RCD technology is limited to 2 axis control, it is relatively simple to implement, with no additional mechanisms or mechanical design complexity, decreasing the risk of failure.

By combining reaction wheels and RCDs, effective control can be retained by using the RCDs to offload momentum from the reaction wheels and prevent reaction wheel saturation. Because spin disturbance torques are the smallest (Orphee et al., 2017), this combined method allows for continuous attitude control of the solar sail. Specifically, the work implements:

1. A novel control strategy of a rigid solar sail with both reaction wheels and RCDs in a Sun synchronous orbit.
2. Two mode control, alternating between Earth pointing for mission operations and Sun pointing for momentum offloading.

This paper highlights some relevant reference frames in Section 2 before outlining the creation of the models in Section 3. The results from the numerical simulations are presented and discussed in Section 4 before concluding in Section 5.

## 2 Reference Frames

The following are some commonly used reference frames that will be referred to throughout the remainder of the paper.

### 2.1 ICRF

The International Celestial Reference Frame (ICRF) is an inertial reference frame with its origin at the centre of the Earth. This frame is used for position and velocity throughout the models. It is treated as equal to the Earth-Centred Inertial (ECI) coordinate system realised at J2000.

## 2.2 LVLH

The Local Vertical, Local Horizontal (LVLH) frame is a rotating frame centred on the spacecraft, useful for describing the orientation of a spacecraft in orbit. Its x axis is defined as the spacecraft position vector from the Earth. Its z axis is defined as pointing normal to the orbital plane. Its y axis completes the right-hand coordinate system and aligns with the direction of the velocity vector for circular orbits.

## 2.3 NED

The North-East-Down (NED) frame is a rotating frame centred on the spacecraft. Its x-axis aligns with true North, its z-axis points towards the interior of the Earth, and its y-axis completes the right-hand coordinate system.

## 2.4 Body Frame

The body coordinate frame is a non-inertial reference frame fixed to the body of the spacecraft. This is shown in Fig. 3 below. The X and Y axis are parallel to the sail plane, and the Z axis is perpendicular to the sail plane.

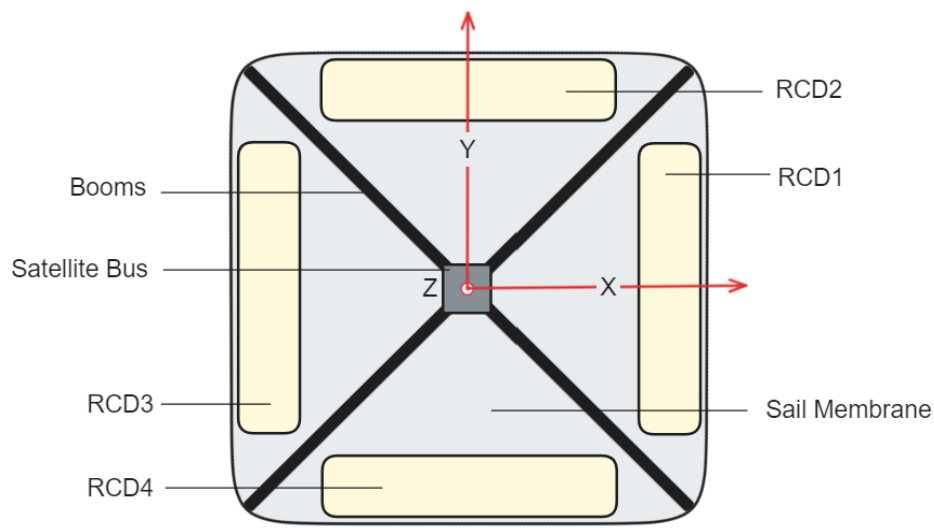

*Figure 3. Solar sail RCD layout and body axes. Z axis perpendicular to sail (out of page). Not to scale.*

# 3 Solar Sail Model

## 3.1 Overview of System

Solar sail models have been developed to test the effectiveness of RCDs at offloading momentum from the reaction wheels and preventing saturation. The solar sails were modelled to be in a Sun Synchronous orbit, typically used for Earth observation (Boldt-Christmas, 2020) and ensures consistent exposure to sunlight. Initial detumbling, and deployment of the solar sail are outside the scope of this paper and are not modelled.

Two models have been developed, henceforth referred to as model A and model B. Model A relies solely on the use of reaction wheels for attitude control, maintaining an Earth pointing orientation. An overview of the control system for model A is shown in Fig. 4 below.

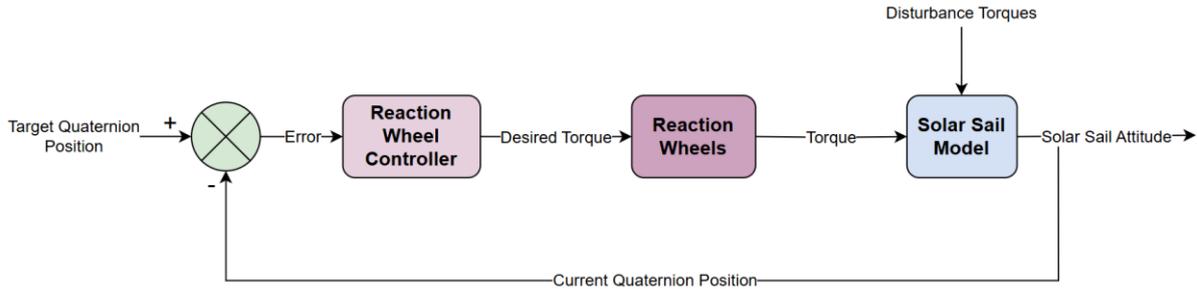

*Figure 4. Model A attitude control system diagram.*

This is a typical feedback loop used for satellite attitude control. Note that the position and rotation of the satellite are assumed to be known exactly, and no sensors or sensor inaccuracies have been modelled for simplicity. The error is calculated from the difference between the current orientation and the target orientation. This error is given to the reaction wheel controller, which calculates the desired torque signal for the reaction wheels. This desired torque is given to the reaction wheel actuators, which generate an actual torque that is dependent on the actuator limitations. This is then applied to the solar sail plant. Meanwhile, disturbance torques also act on the solar sail, which must be countered by the reaction wheels to maintain pointing.

Model B uses RCDs to offload momentum from the reaction wheels to prevent saturation from occurring. Two modes have been implemented in model B: Earth pointing and Sun pointing. The nominal mode is Earth pointing for completing mission objectives like observation, switching to Sun pointing to maximise SRP and utilise the RCDs for momentum offloading. Once offloading is complete, the spacecraft will return to Earth pointing. Note that the RCDs are only used in Sun pointing mode. The control system for model B is shown in Fig. 5 below.

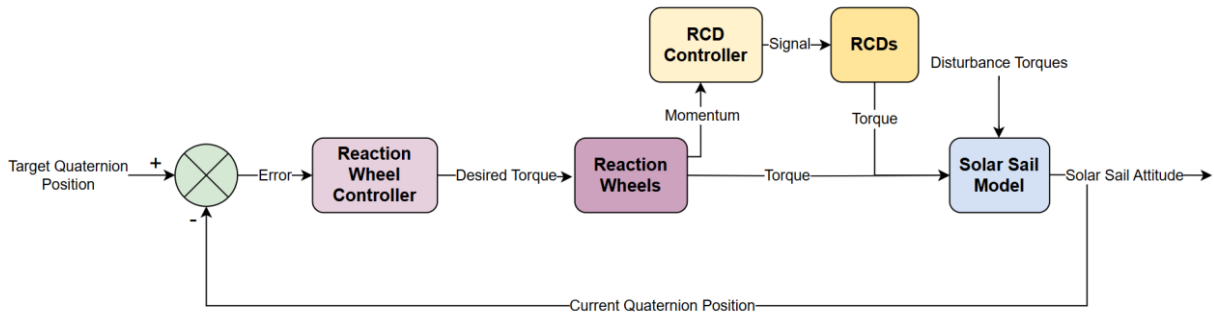

*Figure 5. Model B attitude control system diagram.*

Fig. 5 additionally contains the 'RCD Controller' and the 'RCDs' blocks. The RCD controller commands each of the RCDs to be either ON or OFF to generate the required torque. The torque acts on the solar sail and is counteracted by the reaction wheels, reducing the wheel angular velocity and therefore momentum. The results from model B will be compared with those from model A to evaluate the effectiveness of model B.

### 3.2 Dynamics
#### 3.2.1 Orbital Dynamics

The spacecraft orbital dynamics are governed by the following equation:

$$\boldsymbol{a} = \frac{\boldsymbol{F}_{\text{body}}}{m} + \boldsymbol{a}_{\text{centralbody}} + \boldsymbol{a}_{\text{thirdbody}}, \quad (1)$$

where $\boldsymbol{a}$ is the spacecraft acceleration in ICRF, $\boldsymbol{F}_{\text{body}}$ is the combined atmospheric drag and SRP force acting on the spacecraft body, $m$ is the mass of the spacecraft, $\boldsymbol{a}_{\text{centralbody}}$ is the acceleration from Earth, and $\boldsymbol{a}_{\text{thirdbody}}$ is the acceleration from additional bodies. $\boldsymbol{a}_{\text{centralbody}}$ is determined using the EGM2008 Spherical harmonics gravitational potential model (Pavlis et al., 2012). $\boldsymbol{a}_{\text{thirdbody}}$ is determined using DE432t ephemeris data (Folkner et al., 2014) for the Sun and Moon's locations, assuming point masses. $\boldsymbol{F}_{\text{body}}$ is given by:

$$\boldsymbol{F}_{\text{body}} = \boldsymbol{C}_{\text{body}\to\text{ICRF}}(\boldsymbol{F}_{\text{drag}} + \boldsymbol{F}_{SRP}), \quad (2)$$

where $\boldsymbol{F}_{\text{drag}}$ is the atmospheric drag force in the body frame, and $\boldsymbol{F}_{SRP}$ is the SRP force in the body frame. $\boldsymbol{C}_{\text{body}\to\text{ICRF}}$ is the Direction Cosine Matrix (DCM) converting from the body reference frame to ICRF.

*3.2.2 Attitude Dynamics*

The spacecraft attitude dynamics are governed by the following equation from Euler's rotational equation of motion (Wie, 1998):

$$\boldsymbol{\alpha} = I^{-1}[\boldsymbol{M}_b + \boldsymbol{\tau}_b - \boldsymbol{\omega} \times (I\boldsymbol{\omega})], \quad (3)$$

where $\boldsymbol{\alpha}$ is the angular acceleration of the spacecraft in ICRF, $\boldsymbol{M}_b$ is the torque from actuators (RCDs and reaction wheels), $\boldsymbol{\tau}_b$ is the combined disturbance torques on the spacecraft, $\boldsymbol{\omega}$ is the angular velocity of the spacecraft in ICRF, and $I$ is the spacecraft inertia tensor matrix. $\boldsymbol{\tau}_b$ is given by:

$$\boldsymbol{\tau}_b = \boldsymbol{\tau}_{SRP} + \boldsymbol{\tau}_{\text{drag}} + \boldsymbol{\tau}_M + \boldsymbol{\tau}_{GG}, \quad (4)$$

where $\boldsymbol{\tau}_{SRP}$, $\boldsymbol{\tau}_{\text{drag}}$, $\boldsymbol{\tau}_M$, and $\boldsymbol{\tau}_{GG}$ are the SRP, atmospheric drag, magnetic field, and gravity gradient disturbance torques, respectively.

*3.3 External Forces*

*3.3.1 Solar Radiation Pressure Force*

The SRP force ($\boldsymbol{F}_{SRP}$) is dependent on the reflectivity of the sail surface. Assuming a flat surface, the force is the sum of the forces produced by the light that is specularly reflected ($\boldsymbol{F}_s$), diffusively reflected ($\boldsymbol{F}_d$), and absorbed ($\boldsymbol{F}_a$) as shown below (Farres, 2023):

$$\boldsymbol{F}_{SRP} = \boldsymbol{F}_s + \boldsymbol{F}_d + \boldsymbol{F}_a \quad (5)$$

$\boldsymbol{F}_s$ is represented by (Farres, 2023):

$$\boldsymbol{F}_s = -2\rho_s P_{srp} A \cos^2(\phi)\hat{\boldsymbol{n}}, \quad (6)$$

where $\rho_s$ is the rate of specular reflection, $P_{srp}$ is the solar radiation pressure, $A$ is the reflective area, $\hat{\boldsymbol{n}}$ is the unit vector normal to the solar sail plane, and $\phi$ is the angle between $\hat{\boldsymbol{n}}$ and the Sun spacecraft unit vector ($\hat{\boldsymbol{r}}_s$). $\boldsymbol{F}_d$ is represented by (Farres, 2023):

$$\boldsymbol{F}_d = \rho_d P_{srp} A |\cos(\phi)| \left(\hat{\boldsymbol{r}}_s - \frac{2}{3}\hat{\boldsymbol{n}}\right), \quad (7)$$

where $\rho_d$ is the rate of diffusive reflection. $\boldsymbol{F}_a$ is represented by (Farres, 2023):

$$\boldsymbol{F}_a = \rho_a P_{srp} A |\cos(\phi)|\hat{\boldsymbol{r}}_s, \quad (8)$$

where $\rho_a$ is the rate of absorption. The rates of specular reflection, diffusive reflection and absorption satisfy the condition $\rho_s + \rho_d + \rho_a = 1$. $P_{srp}$ is given by:

$$P_{srp} = \frac{S}{c}, \qquad (9)$$

where $S$ is the solar constant and $c$ is the speed of light. A dual cone eclipse shadow model is used to adjust the SRP force when the spacecraft is eclipsed by the Earth or Moon.

*3.3.2 Atmospheric Drag Force*

The atmospheric drag force ($\boldsymbol{F}_{\text{drag}}$) is calculated using the following (Montenbruck et al., 2002):

$$\boldsymbol{F}_{\text{drag}} = -\frac{1}{2}\rho C_d A_{\text{ram}} |\boldsymbol{v}_{rel}^2| \hat{\boldsymbol{v}}_{rel}, \qquad (10)$$

where $\rho$ is the air density, $C_d$ is the drag coefficient, $A_{\text{ram}}$ is the ram (frontal) area of the satellite, $\boldsymbol{v}_{rel}$ is the velocity relative to the atmosphere, and $\hat{\boldsymbol{v}}_{rel}$ is the unit vector of $\boldsymbol{v}_{rel}$. The NRLMSISE 00 atmosphere model (Picone et al., 2002) was used to generate air density values. $A_{\text{ram}}$ is given by:

$$A_{\text{ram}} = |A\cos(\psi)|, \qquad (11)$$

where $A$ is the area of the solar sail and $\psi$ is the angle between $\hat{\boldsymbol{v}}_{rel}$ and $\hat{\boldsymbol{n}}$. Due to the relative size of the satellite bus, only the solar sail area is considered for the atmospheric drag. $\boldsymbol{v}_{rel}$ is calculated by using the difference between the atmosphere velocity ($\boldsymbol{v}_{atm}$) and the satellite velocity in ICRF, before converting to the body frame. The atmosphere is assumed to rotate with the Earth so $\boldsymbol{v}_{atm}$ is given by (Montenbruck et al., 2002):

$$\boldsymbol{v}_{atm} = \boldsymbol{\omega}_{\text{Earth}} \times \boldsymbol{X}_{ICRF}, \qquad (12)$$

where $\boldsymbol{\omega}_{\text{Earth}}$ is the angular velocity vector of the Earth, and $\boldsymbol{X}_{ICRF}$ is the radial vector of the satellite from Earth in ICRF.

*3.4 Disturbance Torques*

Only four disturbance torques are consequential for a typical Earth orbiting spacecraft (Starin and Eterno, 2011): gravity gradient, magnetic field torques, SRP torque and atmospheric drag torque.

*3.4.1 Solar Radiation Pressure Torque*

SRP torque is the most significant for solar sail applications and is the result of an offset between the centre of pressure and the centre of mass. Inconsistent surface properties or multiple surfaces with different angles result in the centre of pressure changing and a torque being created. The SRP torque ($\boldsymbol{\tau}_{SRP}$) is calculated using (Wertz, 1978):

$$\boldsymbol{\tau}_{SRP} = \boldsymbol{c}_s \times \boldsymbol{F}_{SRP}, \qquad (13)$$

where $\boldsymbol{c}_s$ is the vector from the satellite centre of mass to the SRP centre of pressure.

*3.4.2 Atmospheric Drag Torque*

Atmospheric drag torque works in a comparable manner to SRP torque. When the centre of atmospheric drag pressure (somewhere on the ram area) is not aligned with the centre of mass, a torque is created (Starin and Eterno, 2011). The atmospheric drag torque ($\boldsymbol{\tau}_{\text{drag}}$) is calculated using (Wertz, 1978):

$$\boldsymbol{\tau}_{\text{drag}} = \boldsymbol{c}_a \times \boldsymbol{F}_{\text{drag}}, \tag{14}$$

where $\boldsymbol{c}_a$ is the vector from the satellite centre of mass to the aerodynamic centre of pressure.

*3.4.3 Magnetic Field Torque*

A magnetic field torque is created when the residual magnetic moment of a spacecraft does not align with Earth's magnetic field. The magnetic torque will attempt to align the magnetic fields of the spacecraft and the Earth (Starin and Eterno, 2011). The magnetic field torque ($\boldsymbol{\tau}_M$) is calculated using (Wertz, 1978):

$$\boldsymbol{\tau}_M = \boldsymbol{M} \times \boldsymbol{B}_b, \tag{15}$$

where $\boldsymbol{M}$ is the residual dipole of the satellite, and $\boldsymbol{B}_b$ is the Earth's magnetic field vector in the body frame. The World Magnetic Model 2020 (Chulliat et al., 2020) is used to calculate the Earth's magnetic field vector in the NED reference frame. This is converted to the ECI frame using:

$$\boldsymbol{B}_{ECI} = \boldsymbol{C}_{ECEF \to ECI}(\boldsymbol{C}_{NED \to ECEF} \boldsymbol{B}_{NED}), \tag{16}$$

where $\boldsymbol{B}_{ECI}$, $\boldsymbol{B}_{NED}$ are the Earth's magnetic field vectors in the ECI and NED frames respectively, and $\boldsymbol{C}_{ECEF \to ECI}$, $\boldsymbol{C}_{NED \to ECEF}$ are the DCMs converting from ECEF to ECI frame and from NED to ECEF frame, respectively. The ECI frame is treated as equivalent to ICRF. $\boldsymbol{B}_{ECI}$ is converted to the spacecraft body frame using quaternion rotation (Diebel, 2006):

$$\boldsymbol{B}_b = \boldsymbol{q}^{-1} \otimes \begin{bmatrix} 0 \\ \boldsymbol{B}_{ECI} \end{bmatrix} \otimes \boldsymbol{q}, \tag{17}$$

where $\boldsymbol{q}$ is the satellite attitude quaternion, and $\otimes$ is the operator of a quaternion multiplication.

*3.4.4 Gravity Gradient Torque*

Gravity gradient torque is created when the centre of gravity is not aligned with the centre of mass of the spacecraft (Starin and Eterno, 2011). Since gravity is a function of the altitude of the spacecraft, the force of gravity is not consistently applied over the spacecraft, resulting in a centre of gravity that is offset from the centre of mass. The gravity gradient torque ($\boldsymbol{\tau}_{GG}$) can be calculated using (Wertz, 1978):

$$\boldsymbol{\tau}_{GG} = \frac{3\mu_\oplus}{|\boldsymbol{X}|^3} \widehat{\boldsymbol{X}} \times I\widehat{\boldsymbol{X}}, \tag{18}$$

where $\mu_\oplus$ is the geocentric gravitational constant, $I$ is the inertia tensor of the spacecraft, $\boldsymbol{X}$ is the spacecraft position vector, and $\widehat{\boldsymbol{X}}$ is the unit vector of $\boldsymbol{X}$.

*3.5 Reaction Wheels*
*3.5.1 Theory and Modelling*

Ignoring inefficiencies, the reaction wheel torque is equal in magnitude and opposite in direction to the spacecraft torque so that:

$$\boldsymbol{\tau}_{SC} = -\boldsymbol{\tau}_{RW}, \tag{19}$$

where $\boldsymbol{\tau}_{SC}$ and $\boldsymbol{\tau}_{RW}$ are the spacecraft torque and reaction wheel torque, respectively. The reaction wheel dynamics are modelled as (Wie, 1998):

$$\boldsymbol{\tau}_{RW} = \dot{\boldsymbol{H}}_{RW} = I_{RW}\dot{\boldsymbol{\omega}}_{RW}, \qquad (20)$$

where $\boldsymbol{H}_{RW}$ is the reaction wheel momentum, $I_{RW}$ is the reaction wheel moment of inertia, and $\boldsymbol{\omega}_{RW}$ is the reaction wheel angular velocity. Gyroscopic effects are neglected. The reaction wheel angular velocity is calculated based on the desired torque signal from the reaction wheel controller, using the first order finite difference approximation of the continuous torque expression:

$$\boldsymbol{\omega}_f = \boldsymbol{\omega}_i - \frac{\boldsymbol{\tau}_{des}}{I_{RW}}\Delta t, \qquad (21)$$

where $\boldsymbol{\omega}_f, \boldsymbol{\omega}_i$ are the final and initial angular velocities of the reaction wheels, respectively, $\boldsymbol{\tau}_{des}$ is the desired spacecraft torque, and $\Delta t$ is the timestep. The reaction wheel is assumed to be a rotor with a solid disk. The moment of inertia ($I_{RW}$) is calculated with the following equation:

$$I_{RW} = \frac{1}{2}mR^2, \qquad (22)$$

where $m$ and $R$ are the mass and radius of a singular reaction wheel rotor, respectively. The peak momentum ($h_{\max}$) is used to calculate the maximum angular velocity ($\omega_{\max}$) using:

$$\omega_{\max} = \frac{h_{\max}}{I_{RW}} \qquad (23)$$

The reaction wheel angular velocity is therefore limited between $[-\omega_{\max}, \omega_{\max}]$. The torque output on the spacecraft ($\boldsymbol{\tau}_{SC}$) is calculated from this limited angular velocity vector ($\boldsymbol{\omega}_{lim}$) using:

$$\boldsymbol{\tau}_{SC} = -\frac{\boldsymbol{\omega}_{lim} - \boldsymbol{\omega}_i}{\Delta t}I_{RW} \qquad (24)$$

### 3.5.2 Control Law

Given the desired quaternion spacecraft attitude ($\boldsymbol{q}_{des}$) and the current quaternion spacecraft attitude ($\boldsymbol{q}_t$), the error is given by:

$$\boldsymbol{q}_{\text{error}} = \boldsymbol{q}_{des} - \boldsymbol{q}_t \qquad (25)$$

As $\boldsymbol{q}_{\text{error}}$ is a unit quaternion, its DCM ($\boldsymbol{C}$) is defined as (Diebel, 2006):

$$\boldsymbol{C} = \begin{bmatrix} (q_0^2 + q_1^2 - q_2^2 - q_3^2) & 2(q_1 q_2 + q_0 q_3) & 2(q_1 q_3 - q_0 q_2) \\ 2(q_1 q_2 - q_0 q_3) & (q_0^2 - q_1^2 + q_2^2 - q_3^2) & 2(q_2 q_3 + q_0 q_1) \\ 2(q_1 q_3 + q_0 q_2) & 2(q_2 q_3 - q_0 q_1) & (q_0^2 - q_1^2 - q_2^2 + q_3^2) \end{bmatrix}, \qquad (26)$$

where $q_0$ is the scalar component of $\boldsymbol{q}_{\text{error}}$, and $q_1, q_2, q_3$ are the vector components of $\boldsymbol{q}_{\text{error}}$. The corresponding rotation angles for an XYZ rotation order are defined as (Diebel, 2006):

$$\phi = \text{atan2}(\boldsymbol{C}(2,3), \boldsymbol{C}(3,3)), \qquad (27)$$
$$\theta = -\text{asin}(\boldsymbol{C}(1,3)), \qquad (28)$$
$$\psi = \text{atan2}(\boldsymbol{C}(1,2), \boldsymbol{C}(1,1)), \qquad (29)$$

where $\phi$, $\theta$, and $\psi$ are the rotation angles about the X, Y, and Z axes, respectively. The rotation angle error ($\boldsymbol{e}$) is given by:

$$\boldsymbol{e} = \begin{pmatrix} \phi \\ \theta \\ \psi \end{pmatrix} \tag{30}$$

The control torque output for the PD controller, commonly used for satellite attitude control, is given by:

$$\boldsymbol{\tau}_{des} = K_p \boldsymbol{e} + K_d \dot{\boldsymbol{e}}, \tag{31}$$

where $K_p$, $K_d$ are the proportional and derivative coefficients, respectively.

*3.5.3 Implementation*

Three reaction wheels were modelled in an orthogonal configuration, aligned with the x, y, and z axes of the satellite body. The model is based on RocketLab's RW 0.01 reaction wheel (RocketLab, 2024), which is said to be suitable for smaller CubeSats. The mass and radius of the rotor were assumed to be 90% of the total mass and housing radius of the reaction wheel. $\omega_{\max}$ was set to 4433 rad/s based on the reaction wheel $h_{\max}$. Gain scheduling was implemented for $K_p$ and $K_d$, so that their values vary depending on the error, allowing for better control over a larger range of errors. The lookup table is shown in Table 2 below.

*Table 2. Lookup table for PD controller coefficients*

| $\|\boldsymbol{e}\|^2$ | $K_p$ | $K_d$ |
|---|---|---|
| $7.5 \times 10^{-3}$ | $4 \times 10^{-4}$ | $4 \times 10^{-2}$ |
| $3 \times 10^{-2}$ | $6 \times 10^{-5}$ | $3 \times 10^{-2}$ |

$\|\boldsymbol{e}\|^2$ is the squared norm of the error. Linear interpolation is used to transition between $K_p$ and $K_d$ values. These values were chosen to avoid overshoot for large errors and reduced steady state error when small errors are present. The PD controller output is limited to $\pm 1.0$ mNm as the reaction wheel actuators are limited in their torque output (RocketLab, 2024).

## 3.6 Reflectivity Control Devices
*3.6.1 Theory and Modelling*

IKAROS utilised a total of 72 RCDs across the sail (Tsuda et al., 2011) but only 4 are modelled here for simplicity as shown in Fig. 3. The RCDs provide torques by setting RCDs along one edge of the solar sail to be specular reflection dominant and the opposite edge to be diffusive reflection dominant. Since generally more force is produced for specular reflection than diffusive reflection (Funase et al., 2011), a force imbalance is present, and a torque is generated.

The specular reflection, diffusive reflection and absorption force are modelled using Eq. (6), Eq. (7), and Eq. (8). When an RCD is ON, the reflection/absorption rates ($\rho_s, \rho_d, \rho_a$) change so that specular reflection is dominant, producing a force $\boldsymbol{F}_{ON}$. When an RCD is OFF, the reflection/absorption rates ($\rho_s, \rho_d, \rho_a$) change so that diffusive reflection is dominant, producing a force $\boldsymbol{F}_{OFF}$. When two RCDs are positioned on the solar sail a distance $-d$ and $d$ from the centre of mass, and each side is set to different states, the torque ($\boldsymbol{\tau}_{RCD}$) is given by:

$$\boldsymbol{\tau}_{RCD} = \boldsymbol{d}_1 \times \boldsymbol{F}_{ON}\hat{\boldsymbol{n}} + \boldsymbol{d}_2 \times \boldsymbol{F}_{OFF}\hat{\boldsymbol{n}}, \tag{32}$$

where $\boldsymbol{d_1}$ and $\boldsymbol{d_2}$ are 3x1 vectors representing the distance between each RCD and the spacecraft centre of mass in the body frame. Since the RCDs are located on opposite sides of the sail, $\boldsymbol{d_1} = -\boldsymbol{d_2}$.

*3.6.2 Control Law*

During Sun pointing mode, the RCD controller is given the reaction wheel X and Y angular velocities ($\omega_{RW_X}$ and $\omega_{RW_Y}$). To offload momentum from the reaction wheels, a torque must be produced according to the following:

$$\tau_{RCD_X}(\omega_{RW_X}) = \begin{cases} |\boldsymbol{\tau}_{RCD}| & \omega_{RW_X} > 0 \\ -|\boldsymbol{\tau}_{RCD}| & \omega_{RW_X} < 0 \end{cases}, \quad (33)$$

$$\tau_{RCD_Y}(\omega_{RW_Y}) = \begin{cases} |\boldsymbol{\tau}_{RCD}| & \omega_{RW_Y} > 0 \\ -|\boldsymbol{\tau}_{RCD}| & \omega_{RW_Y} < 0 \end{cases}, \quad (34)$$

where $\tau_{RCD_X}$ and $\tau_{RCD_Y}$ are the torque produced by the RCDs about the X and Y axis, respectively.

*3.6.3 Implementation*

Sun pointing mode is initiated once a reaction wheel speed of 3100 rad/s (~$0.7\omega_{\max}$) has been reached. The reflection/absorption rates for both RCD ON and RCD OFF states are summarised in Table 3 based on values presented by (Kikuchi and Kawaguchi, 2019).

*Table 3. Summary of reflection/absorption rates*

| RCD State | $\rho_s$ | $\rho_d$ | $\rho_a$ |
|---|---|---|---|
| ON | 0.5 | 0.3 | 0.2 |
| OFF | 0.1 | 0.5 | 0.4 |

Due to the configuration of the RCDs shown in Fig. 3, RCD2 and RCD4 are commanded to offload momentum from reaction wheel X (aligned with body x axis), and RCD1 and RCD3 are commanded to offload momentum from reaction wheel Y (aligned with body y axis). Hysteresis control was implemented to avoid rapidly turning the RCDs on and off. If the reaction wheel speed exceeds a magnitude of 200 rad/s, momentum offloading occurs. When the wheel speed is reduced to 100 rad/s, momentum offloading ceases, preventing overshoot from occurring. Because these values are sufficiently spaced, oscillations from rapid transitioning between states is avoided. Four Set-Reset (SR) Flip Flops are used to store the current state of each RCD. Once both reaction wheel X and reaction wheel Y have been slowed below 200 rad/s, the satellite returns to Earth pointing mode and the RCDs are no longer used.

# 4 Numerical Simulation

## 4.1 Model Parameters

The models were implemented in Simulink using the Aerospace Blockset (MathWorks, 2024). A fixed timestep of 30 seconds was used with a Bogacki Shampine (ode3) solver. Sensitivity analysis was conducted to determine the appropriate timestep. Spacecraft properties from the LightSail 2 mission (Mansell et al., 2023) were used and are summarised in Table 4 below.

*Table 4. Spacecraft properties*

| Description | Value | Units |
|---|---|---|
| Mass | 4.93 | kg |

| Moments of Inertia $I_{xx}, I_{yy}, I_{zz}$ | 3.79, 3.79, 7.33 | kg m² |
|---|---|---|
| Products of Inertia $I_{xy}, I_{yz}, I_{xz}$ | $-1.90 \times 10^{-4}, -8.18 \times 10^{-4}, 1.47 \times 10^{-3}$ | kg m² |
| Sail Area | 32 | m² |
| Satellite Bus Dimensions | $0.3 \times 0.1 \times 0.1$ | m |
| Drag Coefficient ($C_d$) | 2.2 | |

The initial satellite orbital parameters were chosen for a 700km Sun synchronous orbit. These are provided in Table 5.

*Table 5. Initial satellite orbital parameters*

| Description | Value | Units |
|---|---|---|
| Epoch | July 11th, 2024 | |
| Semi major Axis | 7 078 140 | m |
| Eccentricity | 0.001 | |
| Inclination | 98.1928 | ° |
| RAAN | 22 | ° |
| Argument of Periapsis | 0 | ° |
| True Anomaly | 0 | ° |

The sail is assumed to have a total reflectivity of 0.9 and a specular reflectivity fraction of 0.89 from optical testing conducted by NASA (Gauvain and Tyler, 2023). This translates to a specular reflection rate ($\rho_s$) of 0.81, diffusive reflection rate ($\rho_d$) of 0.1, and an absorption rate ($\rho_a$) of 0.09 assuming no transmission. Other key parameters are summarised in Table 6.

*Table 6. Additional spacecraft properties*

| Description | Value | Units |
|---|---|---|
| CoM to Aerodynamic CoP vector ($\boldsymbol{c}_a$) | $[0.01, 0.01, -0.15]^T$ | m |
| CoM to SRP CoP vector ($\boldsymbol{c}_s$) | $[0.01, 0.01, -0.15]^T$ | m |
| Spacecraft residual dipole ($\boldsymbol{M}$) | $[5.1962, 5.1962, 5.1962]^T \times 10^{-3}$ | Am² |
| RCD width | 0.5 | m |
| RCD length | 5 | m |

### 4.2 Model A

The results for the Model A spacecraft (no RCDs) are shown below. Fig. 6 displays the rotation angles of the spacecraft in the LVLH frame, Fig. 7 displays the angular velocity of the three reaction wheels.

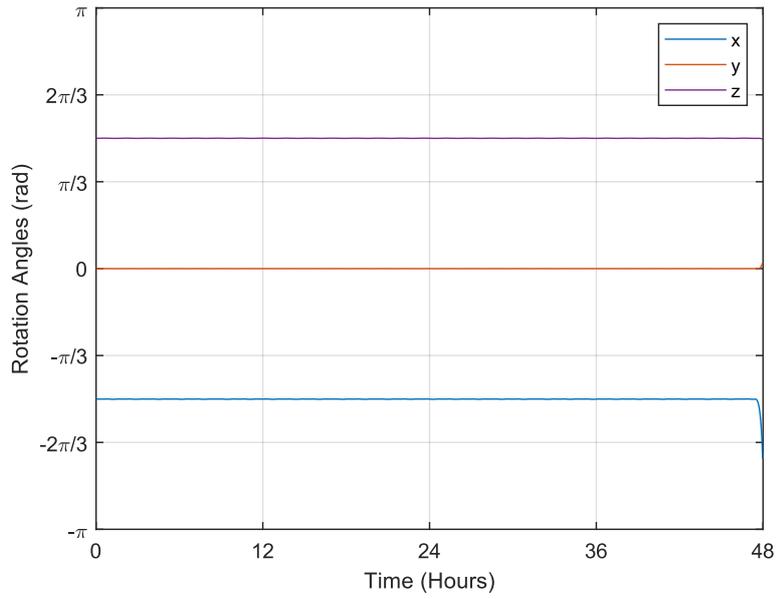

*Figure 6. Model A LVLH to body frame rotation angles about the X axis (blue), Y axis (orange), and Z axis (purple). Rotations are conducted in the order of Z, Y, X.*

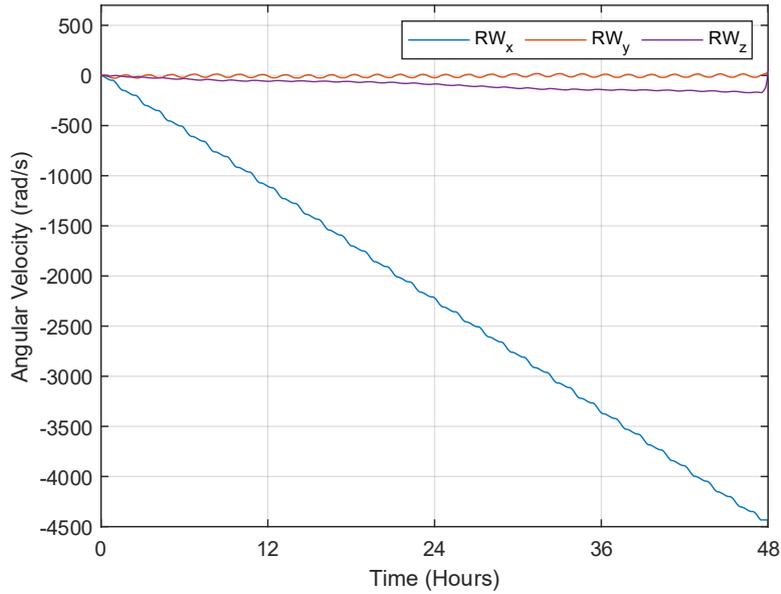

*Figure 7. Model A reaction wheel angular velocities for reaction wheel X (blue), reaction wheel Y (orange), and reaction wheel Z (purple)*

Fig. 6 demonstrates the successful implementation of the control system, while Fig. 7 shows that saturation of reaction wheel X occurs just before 48 hours when an angular velocity of 4433 rad/s is reached. Ignoring the time after saturation occurs, the maximum rotation angle error from Eq. (30) about the X, Y, and Z axis is $1.27 \times 10^{-3}$, $4.38 \times 10^{-4}$, and $6.16 \times 10^{-5}$ radians, respectively. A combined maximum total pointing error of $3.85 \times 10^{-3}$ radians was found using the square root of the sum of the squares of the individual errors. When saturation occurs, 3 axis control of the spacecraft is lost, and the spacecraft starts to spin.

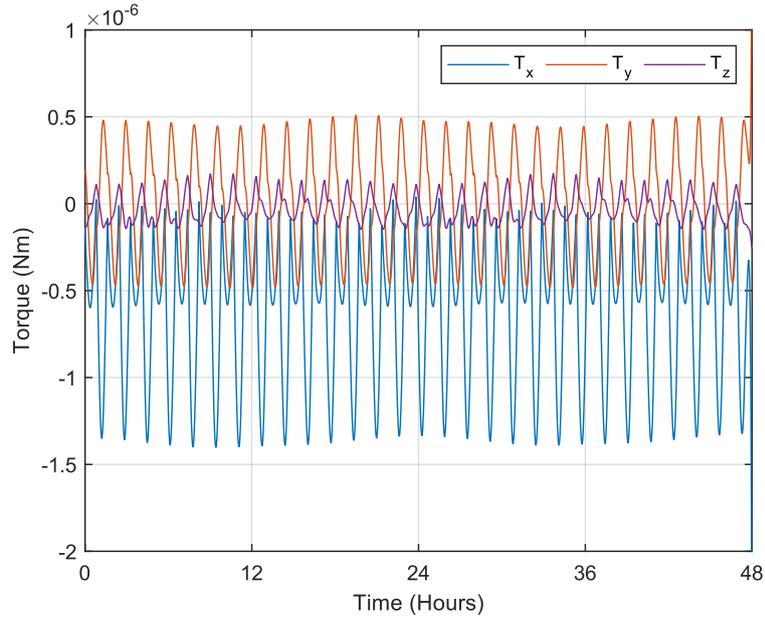

*Figure 8. Model A disturbance torques about X axis (blue), Y axis (orange), and Z axis (purple). Torques are a combination of SRP, atmospheric drag, and magnetic field torques, excluding gravity gradient torque.*

Fig. 8 shows that the X axis experiences the largest disturbance torques, which explains the build-up of momentum in reaction wheel X (Fig. 7). This is largely due to the SRP disturbance torques which measure up to nearly $1.2 \times 10^{-6}$ Nm in the X axis. While the X and Z axes experience bias disturbance torques (net torque), the Y axis experiences mostly zero net torque disturbances resulting from torques balancing out over time. The angular velocity of reaction wheel Y, shown in Fig. 7, therefore does not deviate far from 0. These results demonstrate that using reaction wheels for solar sail attitude control without additional actuators is ineffective due to the bias disturbance torques, which leads to reaction wheel saturation.

### 4.3 Model B

The results for the Model B spacecraft (with RCDs) are shown below. Fig. 9 displays the rotation angles of the spacecraft in the LVLH frame and Fig. 10 displays the angular velocity of the three reaction wheels.

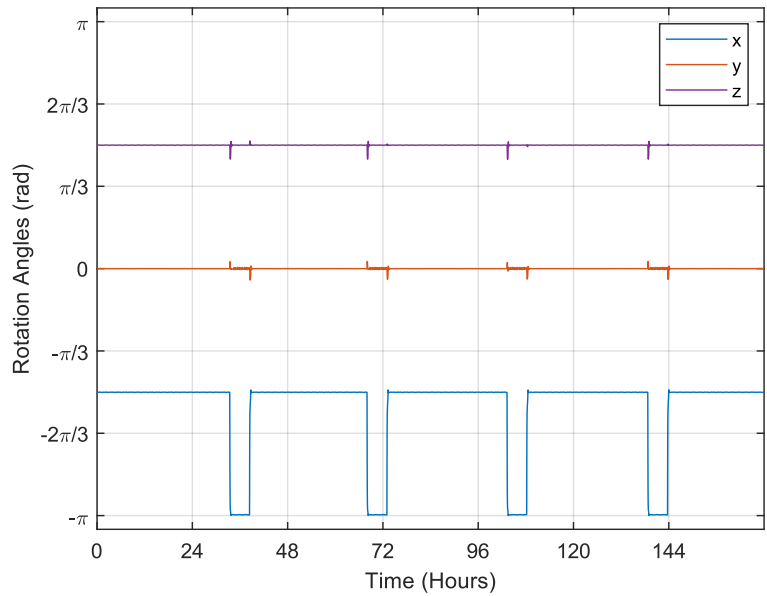

*Figure 9. Model B LVLH to body frame rotation angles about the X axis (blue), Y axis (orange), and Z axis (purple). Rotations are conducted in the order of Z, Y, X*

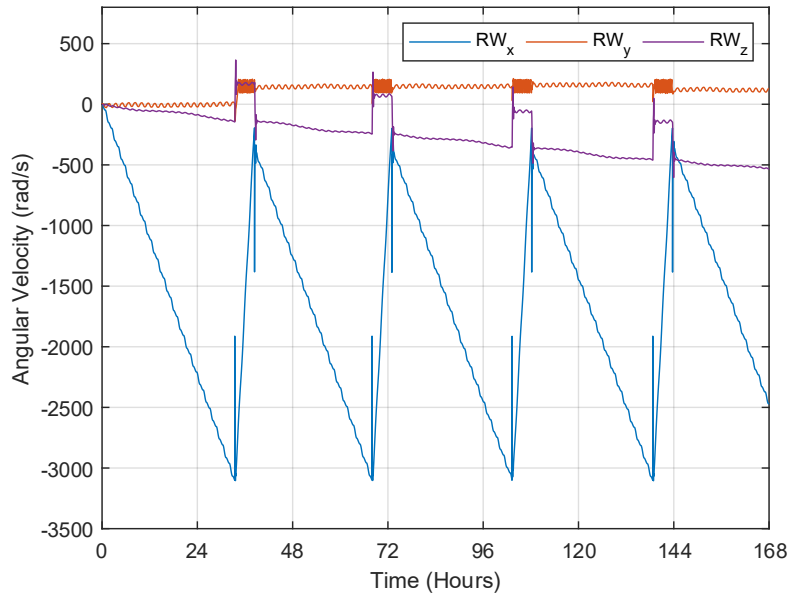

*Figure 10. Model B reaction wheel angular velocities for reaction wheel X (blue), reaction wheel Y (orange), and reaction wheel Z (purple).*

A seven-day period was simulated to assess the momentum offloading capabilities of the RCDs. Fig. 10 shows that momentum offloading occurred four times during the simulation. Sun pointing mode lasted between 4 hours and 58 minutes and 5 hours and 22 minutes running for 11.86% of the total simulated time. Fig. 9 shows that when commanded to be Sun pointing or Earth pointing, rotation was conducted within 11 minutes to within 10° of the target position. While some pointing accuracy is lost during Sun pointing mode, maintaining an exact attitude is not required during this time. The spikes in the reaction wheel X speed (Fig. 10) are the result of large commanded torques, required to rotate the solar sail between Earth pointing and Sun pointing. Fig. 10 shows that momentum was offloaded from both reaction wheel X and reaction wheel Y.

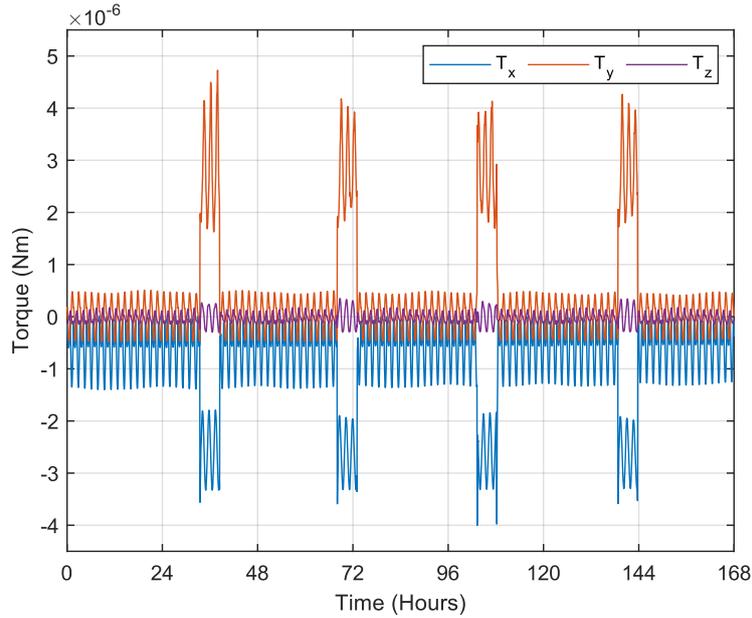

*Figure 11. Model B disturbance torques about X axis (blue), Y axis (orange), and Z axis (purple). Torques are a combination of SRP, atmospheric drag, and magnetic field torques, excluding gravity gradient torque.*

Fig. 11 shows that both X and Y axis disturbance torques increase during momentum offloading. This is primarily due to SRP torque which increases due to a larger Sun facing area. Y axis disturbance torques are greater during Sun pointing because of greater y axis atmospheric drag torque (up to $1.2 \times 10^{-6}$ Nm). The atmospheric drag torque increases in Sun pointing mode as Sun pointing is prioritised over aerodynamics, resulting in a larger ram area of the sail.

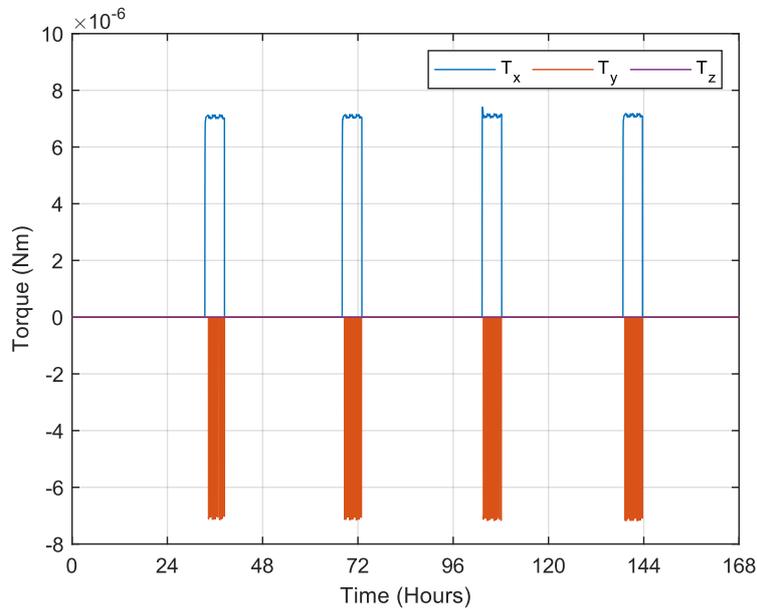

*Figure 12. Model B total RCD torques produced along the body X axis (blue), Y axis (orange), and Z axis (purple).*

The RCDs are therefore used periodically to overcome these disturbances as shown in Fig. 12. A torque of up to $7.4 \times 10^{-6}$ Nm is produced from the RCDs and varies depending on the angle between the solar sail normal and the Sun satellite vector. The y axis RCDs turn on and off rapidly to counter the increase in disturbance torques whilst in Sun pointing mode.

While the RCDs successfully demonstrate their momentum offloading capabilities, preventing saturation of reaction wheel X and Y, reaction wheel Z does increase in angular velocity over time as shown in Fig. 10. This momentum cannot be offloaded using RCDs and will eventually require additional actuators like magnetorquers or thrusters. However, saturation of reaction wheel Z is estimated to take more than 50 days from extrapolation.

*4.4 Sensitivity Analysis*

Due to uncertainty in some of the parameters, a sensitivity analysis was performed to assess how varying key parameters changes the results. These parameters were chosen as they were not found directly in previous literature or expected to alter the results.

*4.4.1 Timestep*

A timestep of 30 seconds was used for the simulations as it was noted that the results were similar to that of 10 seconds and 1 second timesteps for shorter intervals. Timesteps greater than 60 seconds resulted in discrepancies in the angular velocities of the reaction wheels and therefore were not used.

*4.4.2 Residual Dipole*

The residual dipole of the spacecraft was chosen so that $\|M\| = 9 \times 10^{-3}$ Am$^2$ corresponding to the residual dipole of the 3U Space Dart CubeSat (Armstrong et al., 2009). However, (Larson and Wertz, 1999) state that the residual dipole can be from $\|M\| = 0.2$ Am$^2$, depending on the size of the spacecraft. This indicates a large range in values. When using the latter value in the simulations, the magnetic field torque becomes the dominant torque during nominal Earth pointing, and greatly increases momentum in the Z reaction wheel, resulting in saturation after four days.

*4.4.3 Centre of Pressure*

Both the centre of atmospheric drag pressure ($c_a$) and the centre of the SRP ($c_s$) can vary depending on the spacecraft and sail properties. Wie, 2002 considered a solar sail with a centre of mass to centre of SRP offset 0.25% of the sail length. Therefore, for a 5.66 x 5.66m solar sail an offset of magnitude 1.41 cm along the sail plane was used. The model assumes the same value for atmospheric drag centre of pressure. Increasing the centre of pressure offset for both $c_a$ and $c_s$ to 2.42 cm in magnitude increases the disturbance torques on the spacecraft. The SRP disturbance torque is most significant during Sun pointing producing torques up to $5.5 \times 10^{-6}$ Nm. This is the result of larger SRP force produced when fully illuminated by the Sun. As the centre of pressure deviation is heavily dependent on manufacturing and deployment success, this value may vary significantly between missions.

*4.4.4 RCD Reflectivity Rates*

Even with slightly modified reflectivity rates so that the RCDs have lower performance (50% increase in absorption rate when ON), the RCDs can still be utilised to offload momentum. Torques produced are up to $4.7 \times 10^{-6}$ Nm and momentum offloading time is increased (~10 hours).

*4.5 Model Validation*

Spencer et al., 2021 reported that LightSail 2 required daily momentum offloading to prevent saturation of the reaction wheel. This aligns reasonably well with these results as momentum offloading is conducted four times in seven days. However, the LightSail 2 mission required frequent slews and made use of 1 reaction wheel with 3 magnetorquers so a direct comparison is difficult to make.

Kikuchi and Kawaguchi, 2019 provides RCD torque capabilities for different sail lengths between 10 and 300 metres. Using cubic extrapolation, a torque for a 5.66m length solar sail is expected to be $\sim 6 \times 10^{-6}$ Nm in magnitude. This aligns closely to the measured torque of $7.4 \times 10^{-6}$ Nm, indicating that the RCD torque values are reasonable.

*4.6 Assumptions and Limitations*

In all calculations, the spacecraft bus is neglected due to its small size relative to the sail and difficulty to model. This is expected to increase SRP and atmospheric drag torques slightly. The solar sail is assumed to be a rigid body with no wrinkles, tears, or inconsistencies. While the reflection/absorption rates account for some of this, results are likely to vary depending on surface conditions. The atmosphere is assumed to be rotating with the Earth, which is a great simplification, but reasonable in some cases. The reaction wheels are assumed to be perfect actuators, providing instantaneous torque with no inefficiencies. Current implementation is only appropriate for low rotational rates. The position and attitude of the spacecraft are known exactly with results expected to vary due to sensor noise and inaccuracies.

This model is limited in that it only considers a seven-day period of a specific orbit. Performance over longer periods of time and varying orbits is uncertain. Limited real-world data has been used for verification so results may not represent real world outcomes.

*4.7 Implications*

These results indicate the potential application of RCDs for offloading momentum from reaction wheels during solar sailing in a low Earth orbit. While magnetorquers have historically been used for this purpose (Mansell et al., 2020), RCDs could additionally be used to simplify control. This result also indicates additional future applications for RCDs in high orbit or deep space solar sail missions where magnetorquers cannot be used. Since maximum torque from the RCDs is produced when the sail normal and Sun spacecraft vector align, the inclusion of two distinct modes (Earth pointing and Sun pointing) proved to be appropriate. This maximised the time spent in nominal Earth observation by decreasing the momentum offloading period. Reducing atmospheric drag during nominal operations also proved to be significant for reducing disturbance torques and should be considered in future missions.

Further investigation should be conducted into the disturbance torques created from the spacecraft bus. A non-rigid sail should also be modelled as disturbance torques, and non-ideal solar sailing performance is expected. Sensors and more realistic actuators should also be implemented, which could increase pointing error and as a result, increase disturbance torques. Additional disturbances are significant as they may affect momentum offloading if their magnitudes exceed the torque capabilities of the RCDs. Additional orbits and spacecraft properties should also be assessed to determine other possible applications for RCDs. While solar sails with RCDs are identified to be feasible, specific missions or objectives have not been identified and should therefore be investigated.

# 5 Conclusion

Solar sailing uses SRP to generate thrust instead of traditional propulsion systems. This has been successfully demonstrated in JAXA's IKAROS mission and the LightSail 2 mission. IKAROS first demonstrated the capabilities of RCDs to generate torques and perform attitude control. LightSail 2 demonstrated solar sailing in an Earth orbit but suffered from frequent saturation of its reaction wheel due to the large mass moment of inertia associated with solar sails. To prevent this, RCDs are proposed as a method of periodically offloading the reaction wheel momentum, allowing for continuous control throughout the mission.

Two models were developed to assess this: Model A, a solar sail with three orthogonal reaction wheels, and Model B, a solar sail with three orthogonal reaction wheels and RCDs. Orbital and attitude dynamics were modelled alongside SRP, atmospheric drag, magnetic field, and gravity gradient disturbance torques. Numerical simulations were conducted using a custom-built program, developed with Simulink's Aerospace Blockset. A 700km Sun synchronous orbit was chosen for a spacecraft with similar properties to that of LightSail 2. In Model A, saturation of the reaction wheels occurred just before 48 hours. Model B demonstrated by alternating between nominal Earth pointing and Sun pointing (for maximising RCD torque), successful momentum offloading of the X and Y aligned reaction wheels. The Z aligned reaction wheel cannot be offloaded using RCDs and is expected to become saturated after 50 days.

Sensitivity analysis of some key parameters was performed to assess the robustness of the model, and its limitations are stated. RCDs are a viable solution for reaction wheel saturation when used in conjunction with additional actuators and carry little risk due to their flight heritage. Investigation should be conducted into specific missions and applications for this technology to identify where the most value can be added. The model is free to access on GitHub and is available at https://github.com/PierceBoughton/SolarSail-Simulink.

## Acknowledgements

I would like to thank the team at the Geo-Informatics and Space Technology Development Agency (GISTDA) in Thailand for hosting me and providing an introduction into spacecraft attitude and control systems. I would also like to thank Connor Boughton for his feedback and support in developing the models.